\documentclass[Conference]{IEEEtran}
\usepackage{cite}
\usepackage{amsmath,amssymb,amsfonts}
\usepackage{graphicx}
\usepackage{textcomp}
\usepackage{xcolor}
\usepackage{mathtools}
\usepackage{algorithm}
\usepackage{algorithmicx}
\usepackage{algpseudocode}
\usepackage{amsmath}
\usepackage{times}
\usepackage{mathptmx}
\usepackage{multirow}
\usepackage{subfigure}

\newcommand{\ie}{\textit{i}.\textit{e}., }

\begin{document}

\title{A Novel Rate and Channel Control Scheme Based on Data Extraction Rate for LoRa Networks
}

\author{
    \IEEEauthorblockN{Qihao Zhou\IEEEauthorrefmark{1}, Jinyu Xing\IEEEauthorrefmark{1}, Lu Hou\IEEEauthorrefmark{1}, Rongtao Xu\IEEEauthorrefmark{2}, Kan Zheng\IEEEauthorrefmark{1}}\\
\IEEEauthorblockA{
	\IEEEauthorrefmark{1}\textit{Intelligent Computing and Communication (IC$^2$) Laboratory, \\ 
	Key Laboratory of Universal Wireless Communications, Ministry of Education,} \\
	\textit{Beijing University of Posts and Telecommunications}\\
	\IEEEauthorrefmark{2}  \textit{State Key Laboratory of Rail Traffic Control and Safety, Beijing Jiaotong University\\}
	Beijing, China\\
	zqh@bupt.edu.cn}
}

\maketitle

\pagestyle{empty}  
\thispagestyle{empty} 

\begin{abstract}
    Long Range (LoRa) has become one of the most popular Low Power Wide Area (LPWA) technologies, which provides a desirable trade-off among communication range, battery life, and deployment cost. In LoRa networks, several transmission parameters can be allocated to ensure efficient and reliable communication. For example, the configuration of the spreading factor allows tuning the data rate and the transmission distance. However, how to dynamically adjust the setting that minimizes the collision probability while meeting the required communication performance is an open challenge. This paper proposes a novel Data Rate and Channel Control (DRCC) scheme for LoRa networks so as to improve wireless resource utilization and support a massive number of LoRa nodes. The scheme estimates channel conditions based on the short-term Data Extraction Rate (DER), and opportunistically adjusts the spreading factor to adapt the variation of channel conditions. Furthermore, the channel control is carried out to balance the link load of all available channels with the global information of the channel usage, which is able to lower the access collisions under dense deployments. Our experiments demonstrate that the proposed DRCC performs well on improving the reliability and capacity compared with other spreading factor allocation schemes in dense deployment scenarios.
\end{abstract}

\begin{IEEEkeywords}
LoRa, Adaptive Data Rate, Channel Control, Low Power Wide Area, LoRaWAN
\end{IEEEkeywords}

\section{Introduction}
In recent years, the definition of Internet of Things (IoT) has undergone a revolutionary development \cite{Low_power_wide_area}. Low Power Wide Area (LPWA) network paradigm refers to have the ability to provide connectivity to the low-power devices distributed over very large geographical areas \cite{LPWAN_Overview}. Several LPWA technologies and communication standards are emerging, such as Long Range (LoRa) \cite{what_is_LoRaWAN}. LoRa is a physical layer technology, which is developed by Semtech and operates in the unlicensed sub-GHz Industrial Scientific and Medical (ISM) band. Moreover, a non-profit association termed as LoRa Alliance proposes LoRaWAN, which is an open standard defining the communication protocols and system architecture for LoRa networks \cite{what_is_LoRaWAN}. Based on the features of LoRa technology and LoRaWAN protocol, there are two aspects worth considering so as to effectively allocate wireless resources and improve network capacity. The first aspect is that multiple packets with different data rates can be transmitted simultaneously on the same channel. On the other hand, LoRa Gateway has the ability to receive packets simultaneously on multiple channels. Therefore, how to properly allocate data rate and control available channels has a significant impact on the performance of LoRa networks.

Although most of the existing studies focus on evaluating the performance of LoRa network as well as coverage and scalability \cite{do_lora_scale,understanding_limits,performance_lora,iotcloud,Soft_defined}, the impact of Adaptive Data Rate (ADR) scheme is not thoroughly investigated. Moreover, several related works propose some methods to control the transmission parameters with different objectives \cite{adaptive_configuration,EXPLoRa,lora_transmission,power_sf_control,Energy-Efficient}. To the best of our knowledge, the existing works focus on either data rate allocation or channel selection control.

In the paper, the proposed Data Rate and Channel Control (DRCC) scheme is responsible for adjusting the data rate of LoRa nodes and balancing the link load. LoRa supports multiple spreading factors to coordinate the trade-off between communication range and data rate. Therefore, the scheme is designed to adjust the spreading factors of LoRa nodes to adapt the variation of channel conditions. In order to detect the deterioration as soon as possible, the channel condition is estimated by keeping track of the short-term Data Extraction Rate (DER). DER is defined as the ratio of successfully received messages by LoRa Gateway to transmitted messages by the LoRa node within a estimation window \cite{do_lora_scale}. Furthermore, the proposed scheme is able to manage the channel configuration so as to improve the efficiency of radio resources. DRCC takes advantage of the global knowledge of channel usage to ensure that the number of LoRa nodes using the same data rate on each channel is the same. As a result, each channel is fully utilized to attenuate the access collisions between packets concurrently transmitted. Finally, extensive experiments are carried out to compare the impact of using various spreading factor allocation schemes on the performance of LoRa network. Simulation results demonstrate that DRCC outperforms the basic ADR scheme in terms of reliability and capacity of LoRa networks under dense deployments. 

The rest of the paper is organized as follows. Section II discusses the related works on spreading factor allocation schemes. Section III provides an overview on LoRa system model. Section IV describes DRCC in detail. Section V demonstrates and analyzes the experimental results. Finally, Section VI concludes the paper.

\section{Related Work}
Transmission parameter configuration mechanisms such as ADR scheme need to be executed on both LoRa node and LoRa network server. Taking into account low power consumption, the mechanism running on LoRa node should be as simple as possible and has been detailed in LoRaWAN. However, LoRa network server is responsible for the complex management mechanism, which can be carefully designed to improve network performance. Therefore, the discussed related works focus on server-side mechanism. In addition, the mechanism running on LoRa node is in accordance with the definition of LoRaWAN 1.1 specification \cite{LoRaWAN_specification}.

The basic ADR scheme provided by LoRaWAN estimates channel conditions using the maximum value of the received signal-to-noise ratio (SNR) in several recent packets \cite{LoRaWAN_specification}. When the variance of the channel is low, using ADR scheme significantly reduces the interference and increases the system capacity compared with using the static data rate \cite{do_lora_scale,adaptive_configuration,SMDP}. However, the scheme may also have potential drawbacks. First, SNR measurements are determined by different models of LoRa Gateway. Therefore, the value of SNR is inaccurate as a result of hardware calibration and interfering transmissions. Second, selecting the maximum SNR in the last 20 packets is not an desirable method. Because there may be a long time interval between consecutive packets for some IoT applications. The antiquated SNR information is not able to accurately estimate the channel condition for the next uplink packet. Third, the scheme only considers the link of single node to decide whether to adjust transmission parameters. If massive LoRa nodes are densely distributed near LoRa Gateway, it will cause most of nodes using the fastest data rate. With the number of LoRa nodes using the same data rate increases, the possibility of collisions also increases dramatically.

Moreover, a lot of researchers propose various approaches to allocate transmission parameters with different objectives. Most of the approaches utilize SNR or RSSI information to control transmission power and spreading factor. The authors in \cite{adaptive_configuration} slightly modify the basic ADR scheme. The maximum operation in the SNR of several recent packets is replaced with the average function. In \cite{EXPLoRa}, EXPLoRa-SF selects spreading factors based on the total number of connected nodes and EXPLoRa-AT equalizes the time-on-air of packets transmitted at the different spreading factors. In addition, the authors in \cite{lora_transmission} propose an link probing regime to select transmission parameters in order to achieve lower energy consumption. In \cite{power_sf_control}, the authors present a scheme to optimize the packet error rate fairness to avoid near-far problems.
\section{System Model}

\subsection{Overview on LoRa Network}\label{AA}
\begin{figure}[]
    \includegraphics[width=0.49\textwidth]{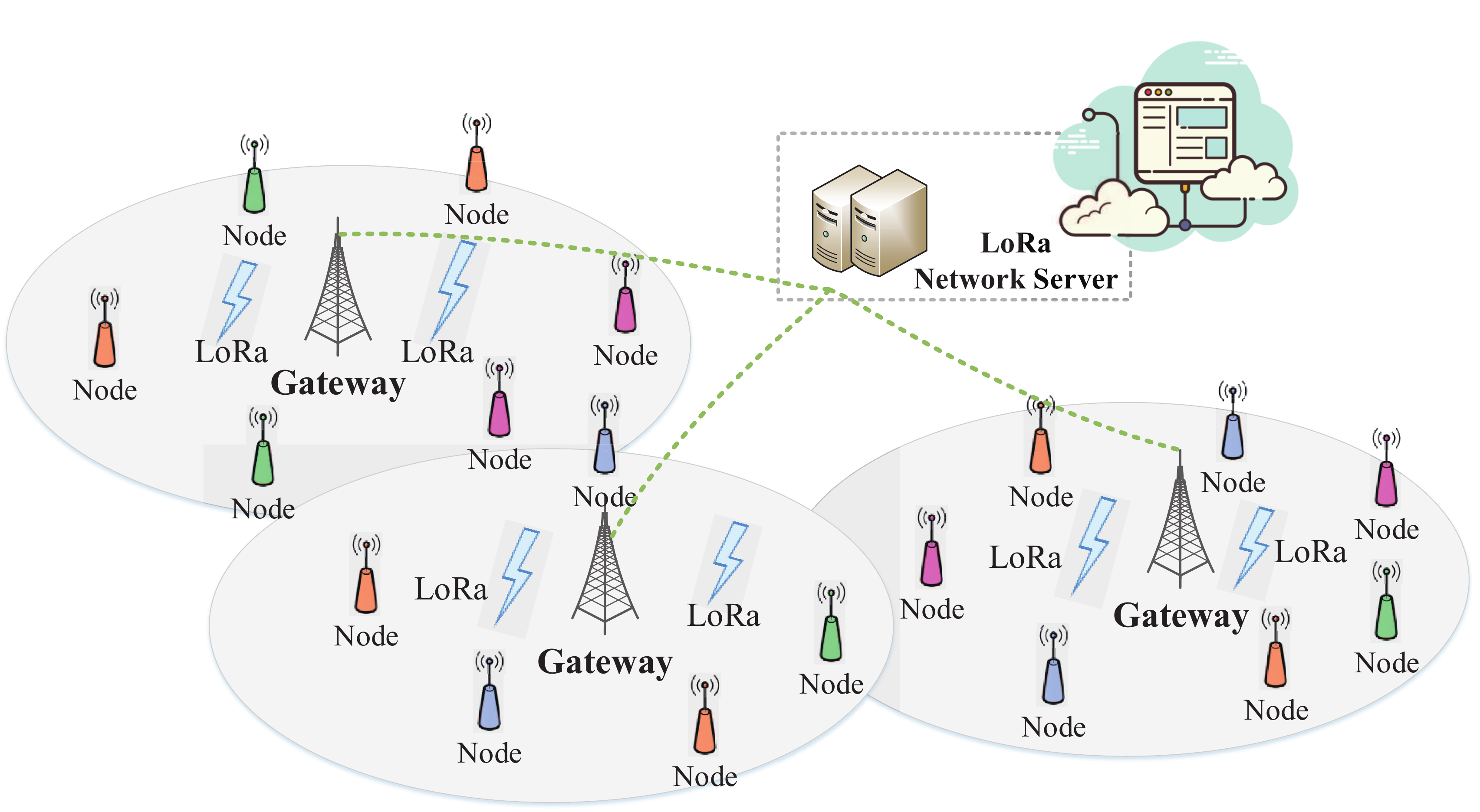}
    \caption{The architecture of LoRa network.}
	\label{architecture_LoRa}
	\vspace{-1em}
\end{figure}
LoRa is a wireless modulation based on Chirp Spread Spectrum (CSS), which enables the energy-constraint devices distributed over wide areas to establish affordable connectivity with the Internet. LoRaWAN defines the network protocol and system architecture \cite{what_is_LoRaWAN}. The two components become the cornerstone of LoRa network. As shown in Fig.\ref{architecture_LoRa}, the architecture of LoRa network uses the star topology.

In order to make the long range star network viable, there are two important characteristics in LoRa network. Firstly, LoRa Gateway uses a multi-channel transceiver so that multiple packets can be received simultaneously on different channels. Secondly, LoRa technique is based on chirp spread spectrum modulation and supports multiple spreading factors \cite{what_is_LoRaWAN}. As long as LoRa nodes use different spreading factors, the signal that transmits the packet is orthogonal and can be decoded concurrently. LoRa Gateway takes advantage of these properties to receive packets with different data rate on the same channel simultaneously. On the other hand, the configuration of the spreading factor allows tuning range and data rate. The higher spreading factor corresponds to the higher coverage distance. However, it implies the slower data rate, the longer transmission time and the higher energy consumption. Therefore, when LoRa nodes use the high spreading factor, the communication channel is occupied for a long time and the possibility of collisions increases.

\subsection{LoRa MAC Control Method}
For network administration, MAC commands are defined by LoRaWAN and the basis for implementing the modification of node transmission parameters. They are exchanged between LoRa network server and the MAC layer on LoRa node \cite{LoRaWAN_specification}. In ADR schemes, a pair of MAC commands named as LinkADRReq and LinkADRAns are used by server and node respectively.

\begin{figure}[]
    \centering
    \subfigure[]{
    \begin{minipage}[a]{0.48\textwidth}
    \includegraphics[width=1\textwidth]{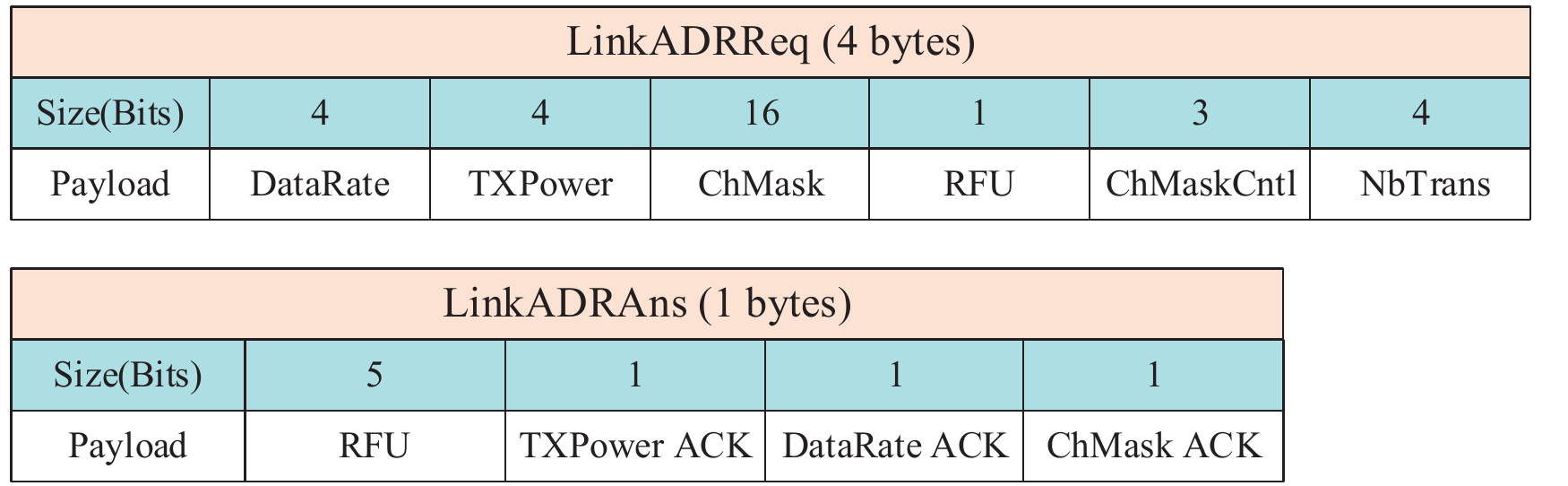} \\
    \end{minipage}
    }
    \subfigure[]{
    \begin{minipage}[b]{0.48\textwidth}
    \includegraphics[width=1\textwidth]{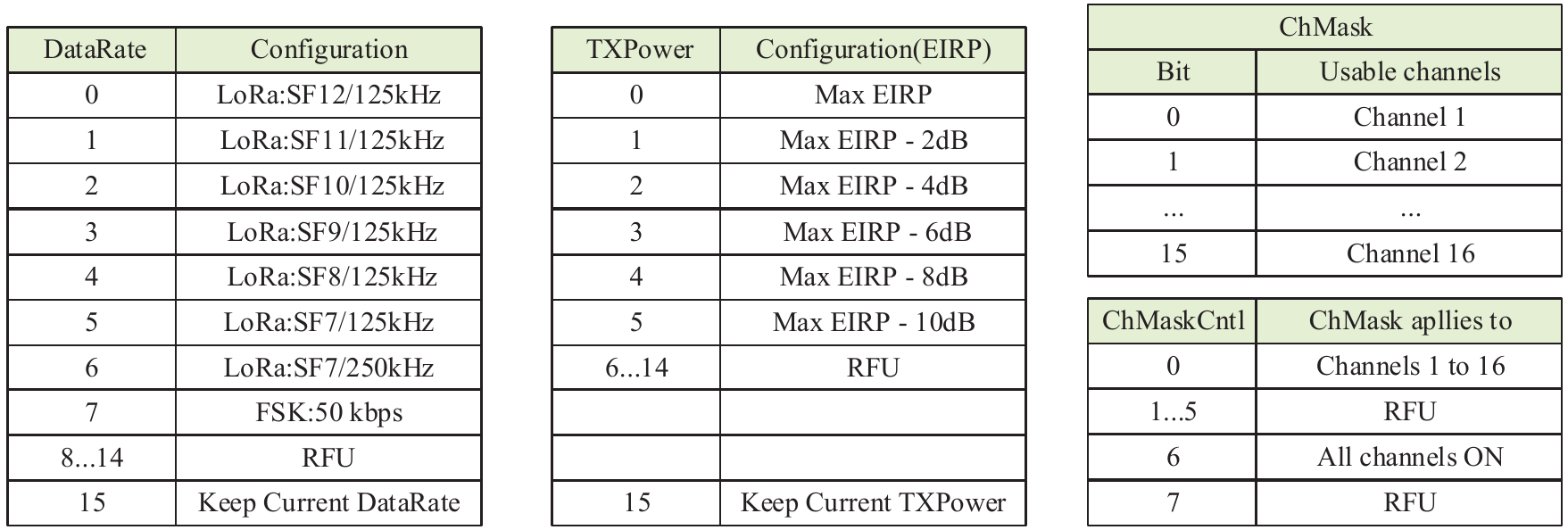} \\
    \end{minipage}
    }
	\caption{(a) The formats of LinkADRReq and LinkADRAns. (b) The optional parameters in LinkADRReq at EU 868MHz ISM Band.}
	\label{linkADR_LoRa}
	\vspace{-1em}
\end{figure}

The formats of LinkADRReq and LinkADRAns are shown in Fig.\ref{linkADR_LoRa}(a) \cite{LoRaWAN_specification}. The LinkADRReq includes five adjustable communication parameters. DataRate and Txpower control data rate and transmission power of LoRa node respectively. ChMask and ChMaskCntl manage the available channels according to  operating frequency band. NbTrans is the maximum number of transmissions for each uplink packet. We present an example about these parameters used at 868 $MHz$ ISM Band. As shown in Fig.\ref{linkADR_LoRa}(b), there are eight levels for DataRate selection and six levels for TXPower. RFU is reserved for future usage. In addition, the maximum Equivalent Isotropically Radiated Power (EIPR) is 16 $dBm$ and the total number of available channels is sixteen. When receiving the LinkADRReq, the node must answer the LinkADRAns to acknowledge whether has successfully configured the parameters. Parameter allocation schemes basically start with the adjustment of data rate, transmission power and available channel.

\section{Design on DRCC}

In the section, the well-design DRCC scheme is elaborated in details aiming at improving LoRa network throughput and achieving the following goals:

\begin{itemize}
\item Quickly track of random loss. With a massive number of LoRa nodes, random loss due to the variation of channel conditions and the interference between packets is unavoidable. The channel condition estimated by SNR/RSSI is antiquated and not accurate. The scheme is expected to use a reasonable approach for adjusting the data rate and properly respond to channel condition deterioration.
\item Attenuate access collision in dense deployment scenarios. LoRa node has multiple channels to choose for uplink and LoRa Gateway has the ability to receive packets simultaneously on different channels. These play an important role to improve the capacity of LoRa network. Therefore, the scheme should take account of these characteristics and make full use of the radio resources.
\end{itemize}

For the sake of achieving the above goals, DRCC combines both data rate allocation and channel selection control. DRCC choses the short-term DER as the indicator to estimate channel condition, and determines whether to adjust the data rate with multiple thresholds. Moreover, the global information of the channel usage is used to evenly allocate LoRa nodes which are using the same data rate over all available channels.

\subsection{DRCC}
The data rate control mechanism in DRCC is used to adjust the spread factor by the link-layer information of the short-term DER. The short-term DER is defined as the ratio of successfully received messages to transmitted messages in a short period \cite{do_lora_scale}. The details of mechanism are described in Algorithm \ref{algorithm_iteration} (line 7-17). There are four important parameters associated with the adjustment strategy, \ie the short-term DER, Minimum Tolerable Success threshold (MTS), Proposed Rate Increase threshold (PRI), and Saturated Quantity threshold (SQI).
\begin{algorithm}
	\caption{DRCC}
	\label{algorithm_iteration}
	\begin{algorithmic}
		\Require ~~\\
		$SF$: The spreading factor used by the node\\
		$RSSI$: The RSSI of the latest packet\\
		$SFGroup$ [SF]: The number of nodes using the specific SF\\
		$ChCtrl$ [SF][CH]: The number of nodes using the specific SF on channel CH
		\Ensure ~~\\
		$SF$: The spreading factor used for next transmission\\
		$Ch$: Available channel for next transmission
	\end{algorithmic}
	\begin{algorithmic}[1]
		\State \# Allocate spreading factor
		\State \# The default SF vector
		\State $SF_{set} \Leftarrow$ [7,8,9,…,12]
		\State \# All available channels vector
		\State $Ch_{set} \Leftarrow$ [1,2,3,…,8]
		\State $P$: The short-term DER
		\If {$P <$ MTS and $SFGroup$ [$SF$+1] $<$ SQI [$SF$+1]}
		\State $SF \Leftarrow$ $SF$+1
		\EndIf
		\If {$P >$ PRI and $SFGroup$ [$SF$-1] $<$ SQI [$SF$-1]}
		\If {$RSSI >$ Sensitivity[$SF$-1]}
		\State $SF \Leftarrow$ $SF$-1
		\Else
		\State Return $SF$
		\EndIf
		\EndIf
		\State Return $SF$
		\State \# Balance channel load
		\If {$SF$ is modified}
		\State $SFb$, $SFa$: SF before and after adjustment
		\State \#Delete records in $SFGroup$ and $ChCtrl$
		\State $MinNumChIndex \Leftarrow$ Index[min($ChCtrl$[$SFa$])]
		\State $Ch \Leftarrow$ $Ch_{set}$[$MinNumChIndex$]
		\State \#Add records in $SFGroup$ and $ChCtrl$
		\EndIf
		\State Return $Ch$
	\end{algorithmic}
\end{algorithm}

In order to obtain the short-term DER, the estimation window is defined in advance. It is the fixed number of received packets by LoRa network server. The size of estimation window is not directly related to time. For different application, the time interval sending consecutive packets is different. If the estimation window is set to 10, LoRa network server uses the last 10 packets of the node. The short-term DER is calculated as:
\begin{equation}
    P =	\frac {R} {T} \label{eq1},
\end{equation}
\noindent where $R$ is the number of the received packets and equal to the size of estimation window. $T$ is total number of the packets transmitted by the node within the window. $T$ can be calculated by Frame Counter (FCnt), which must be incremented except retransmission and included in the payload of each uplink packet. Therefore, the total number of packets sent by the node can be obtained by the difference between the latest FCnt and the oldest FCnt in the estimation window. It is worth mentioning that the server is responsible for removing duplicate packets.

In the following, the three thresholds, \ie MTS, PRI and SQI, are described. The MTS and PRI can be set according to  requirements respectively, such as 40\% and 80\%. SQI is referred as $\Gamma$ and the coefficient $\alpha(s)$ is derived from \cite{power_sf_control}. They can be calculated as:
\begin{equation}
    \Gamma(s)= \alpha(s) * N \quad \forall s \in SFs \label{eq3},
\end{equation}
\begin{equation}
    \alpha(s)=\frac{s}{2^{s}} / \sum_{i=7}^{12}\frac{i}{2^{i}} \quad \forall s \in SFs \label{eq4},
\end{equation}
\noindent where $N$ is the total number of LoRa nodes and $\alpha(s)$ presents the reference percentage of nodes using the  Spreading Factor $SF$. The configuration of $SF$ allows from 7 to 12.The sum of all coefficients should be unity $ \sum_{s=7}^{12}\alpha(s)=1$. The mechanism adjusts $SF$ in accordance to the workflow. When a new uplink packet arrives, LoRa network server calculates $P$ of the node in the estimation window. If $P$ at current $SF$ is below MTS and the number of nodes using $SF+1$ is below SQI, it implies that the channel condition using $SF$ is not optimistic and more nodes can be accommodated at higher $SF$. Therefore, the $SF$ can be increased one level in order to improve transmission stability. On the other hand, if $P$ exceeds PRI, it implies that the channel is idle and the $SF$ can be opportunistically decreased to reduce the occupation of radio resources. Decreasing $SF$ requires careful consideration because it also reduces the transmission distance. The short transmission distance may prevent nodes from communicating with any LoRa Gateway. Therefore, the check is necessary to ensure that the RSSI of latest packet is higher than the receiver sensitivity using $SF-1$. The sensitivity of receiver is shown in Table \ref{receiver_sensitivity} \cite{SX1276}.

\begin{table}[]
    \caption{Receiver Sensitivity (in $dBm$) for Different Spreading Factors}
    \begin{center}
        \begin{tabular}{c|lcl}
            \hline
            \multirow{2}{*}{\textbf{Spreading Factor}} & \multicolumn{3}{c}{\textbf{Bandwidth (kHz)}} \\ \cline{2-4} 
                                                       & 125           & 250           & 500          \\ \hline
            7                                          & -123          & -120          & -116         \\
            8                                          & -126          & -123          & -119         \\
            9                                          & -129          & -125          & -122         \\
            10                                         & -132          & -128          & -125         \\
            11                                         & -133          & -130          & -128         \\
            12                                         & -136          & -133          & -130         \\ \hline
        \end{tabular}
    \end{center}
    \label{receiver_sensitivity}
    \vspace{-2em}
\end{table}

In order to mitigate collisions between nodes using the same data rate, the channel load balancing mechanism is proposed in DRCC. The mechanism consists of two components, \ie the initialization process and the rebalancing process. 

\begin{figure}[b]
	\vspace{-2em}
	\centering
	\includegraphics[width=0.50\textwidth]{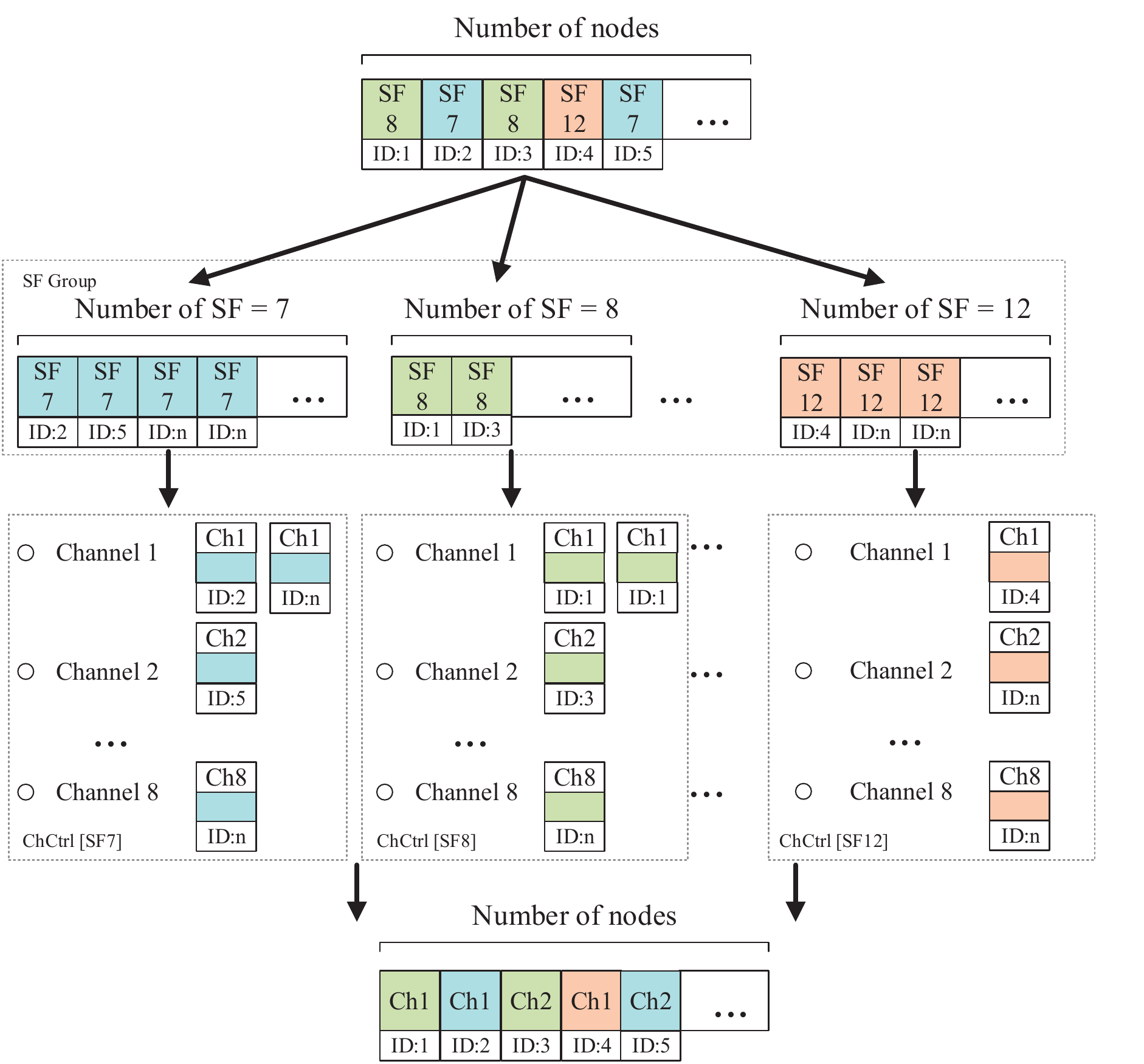}
	\caption{The initialization process of channel load balancing mechanism.}
	\label{ChannelControl}
\end{figure}

The initialization process is carried out when LoRa network server is first implemented. The specific details are shown in Fig.\ref{ChannelControl}. The server collects the transmission information of all access nodes and groups nodes according to $SF$. Then the nodes in the same group are evenly allocated to all available channels in order to achieve initial link load balancing. First, we consider a general situation that $G$ gateways are deployed and $L$ LoRa nodes are able to access the network. In addition, $R$ is the number of default spread factors and $C$ is the number of available channels. Then we define $SF_{nodes}$ which collects the $SF$ of all nodes, a $[1 \times R]$ vector as $SF_{set}$ contains all usable spreading factors and a $[1 \times C]$ vector as $C_{set}$ stores all available channels. In Algorithm \ref{algorithm_iteration}, the $R$ is equal to 6 and the configuration range of $SF_{set}$ is from 7 to 12.
LoRa nodes using the same $SF$ are assigned to a group referred as $SFGroup [SF]$ and the number of groups is determined by $R$. After configuring these parameters, $N$ is determine by the number of LoRa nodes in $SFGroup [SF]$. Next, $N$ LoRa nodes are averagely divided by the available channels $C_{set}$ and $n$ is the number of nodes allocated per channel. If the number of available channels is 8, we know that $n=N/8$. The first $n$ nodes is allocated to use channel 1, and continue the process until the last subset of $n$ nodes are allocated to use the last channel. All groups deal with the above procedure and the channel allocation for each node is stored in $ChCtrl$. Finally, the load on each channel is briefly balanced if the $SF$ of LoRa nodes does not change.

When the $SF$ of a node is modified or a new node is connected, the load balancing of channels is broken. The node needs to be reconfigured with global information of the current channel usage. The rebalancing process is executed which is presented in Algorithm \ref{algorithm_iteration} (line 18-26). $SFb$ and $SFa$ are the spreading factors before and after adjustment. We choose the most suitable channel which has the fewest number of nodes using the $SFa$. When the transmission interval and length of packets are the same, the number of nodes represents the degree of channel load. Therefore, the LoRa node is allocated to use the selected channel. In addition, updating the channel allocation record of the node is essential since it is able to guarantee the timeliness and correctness of each node information.

\subsection{Implement DRCC in LoRaWAN}
The implement of our proposed scheme in a real LoRa network is flexible and straightforward. This scheme can be perfectly integrated with the LoRaWAN specification. The configuration of spreading factor is based on the short-term DER and the channel control is performed with the global knowledge of the channel usage. This information required by the scheme is obtainable by LoRa network server. In addition, the established MAC control method can meet the requirements on the adjustment of data rate and channel. When the optimized communication parameters are computed by the scheme, the downlink packet with LinkADRReq is transferred to LoRa node.

\section{Experimental Results and Analysis}
In order to evaluate the proposed scheme, DRCC has been implemented by extending the LoRaSim \cite{do_lora_scale}. LoRaSim is an open-source discrete-event LoRa simulator based on SimPy \cite{SimPy}. In order to compare the performance with different ADR approaches, we also develop the basic ADR and the spreading factor control method proposed in \cite{power_sf_control}. We performed extensive experiments for different scenarios and the various types of collision behavior are considered \cite{do_lora_scale}. The main parameters of experiments are in Table \ref{simulation_settings}.

\begin{table}[!b]
	\vspace{-2em}
	\caption{Simulation Parameters Settings}
	\begin{center}
		\begin{tabular}{lc}
			\hline
			\multicolumn{1}{c}{\textbf{Parameters}} & \textbf{Values}                                                                                   \\ \hline
			Number of Channels                      & 8                                                                                                 \\
			Channel Centre Frequency (Mhz)          & \begin{tabular}[c]{@{}c@{}}{[}868.1,868.3,868.5,868.7, \\ 868.9,869.1,869.3,869.5{]}\end{tabular} \\
			Bandwidth (kHz)                         & 125                                                                                               \\
			Coding Rate                             & 4/5                                                                                               \\
			Transmit Power (dBm)                    & 14                                                                                                \\
			Spreading Factor                        & 7 to 12                                                                                           \\
			Programmed Preamble                     & 8                                                                                                 \\
			Payload Length (Byte)                   & 20                                                                                                \\
			Proposed Rate Increase threshold        & 80\%                                                                                              \\
			Minimum Tolerable Success threshold     & 40\%                                                                                              \\ \hline
		\end{tabular}
	\end{center}
	\label{simulation_settings}
\end{table}

In the evaluation experiments, the log-distance path loss model \cite{path_loss} is applied, which is commonly used to model dense deployment scenarios. The path loss based on the distance between the node and the gateway can be described as:
\begin{equation}
    L_{pl}(d)=\overline{L_{pl}}(d_{0})+10 \gamma \log{\frac{d}{d_{0}}} + X_{\sigma} \label{eq5},
\end{equation}

\noindent where $L_{pl}(d)$ is the path loss in $dB$, $\overline{L_{pl}}(d_{0})$ is the mean path loss at the reference distance $d_{0}$, $\gamma$ is the path loss exponent and $X_{\sigma} \sim N(0,\sigma^2)$ is the normal distribution with zero mean and $\sigma^2$ variance accounts for shadowing. In the experimental environment, we determine the value of these parameters that $d_{0}$ is 40 meters, $\overline{L_{pl}}(d_{0})$ is 127.41 $dB$, $\gamma$ is 2.08 and $\sigma$ is 0.
\begin{figure}[]
	\centering
	\includegraphics[width=0.50\textwidth]{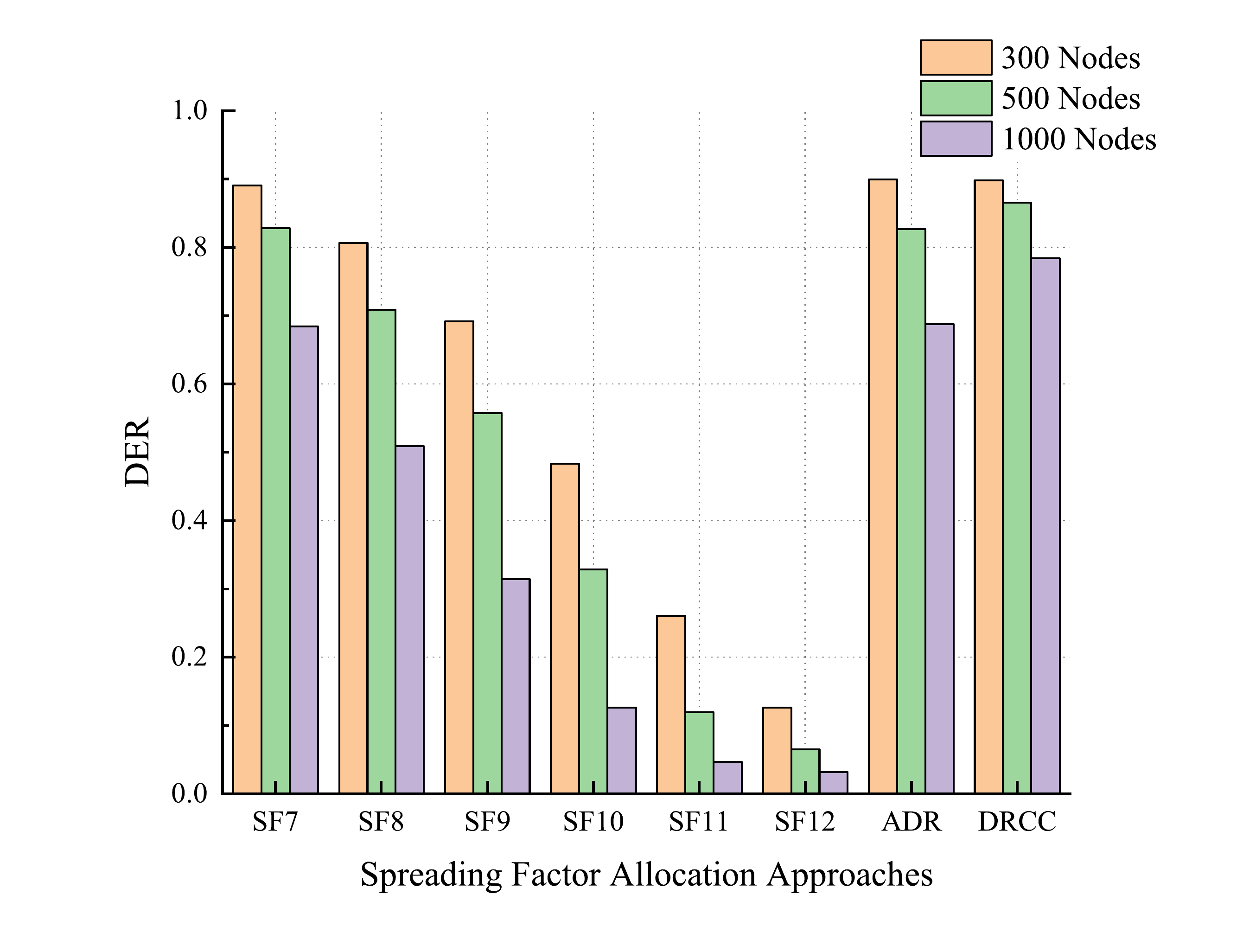}
	\caption{Impact of approaches on DER in LoRa network.}
	\label{compare}
	\vspace{-1em}
\end{figure}

The first experiment aims to demonstrate the impact of different spreading factor allocation methods on system performance. In order to make all nodes visible to the gateway no matter what spreading factor is used, these nodes are distributed in a circle with a radius of 50 meters around a LoRa Gateway. Each node sends an uplink packet with a period of about 30 seconds and eight communication channels can be selected. As shown in Fig.\ref{compare}, there are eight approaches to allocate the data rate of LoRa node. The first six approaches of them use conservative transmission settings, so that all the nodes are running at a static data rate. With the increase of the spreading factor, the time to transfer a packet is greatly increased as well as the possibility of collisions. ADR has basically the same performance as using static SF7. This is because the value of SNR computed by LoRa Gateway is high when LoRa nodes are densely populated in a smaller range. Most of nodes under the basic ADR schemeare are assigned using the fastest rate, \ie SF=7. As the number of nodes increases, the collisions are inevitable when packets are transmitted concurrently. DRCC can dynamically adjust nodes to use other data rates if serious collisions occurs. Therefore, most of collisions caused by dense deployments are weakened.

\begin{figure}[]
	\centering
	\includegraphics[width=0.50\textwidth]{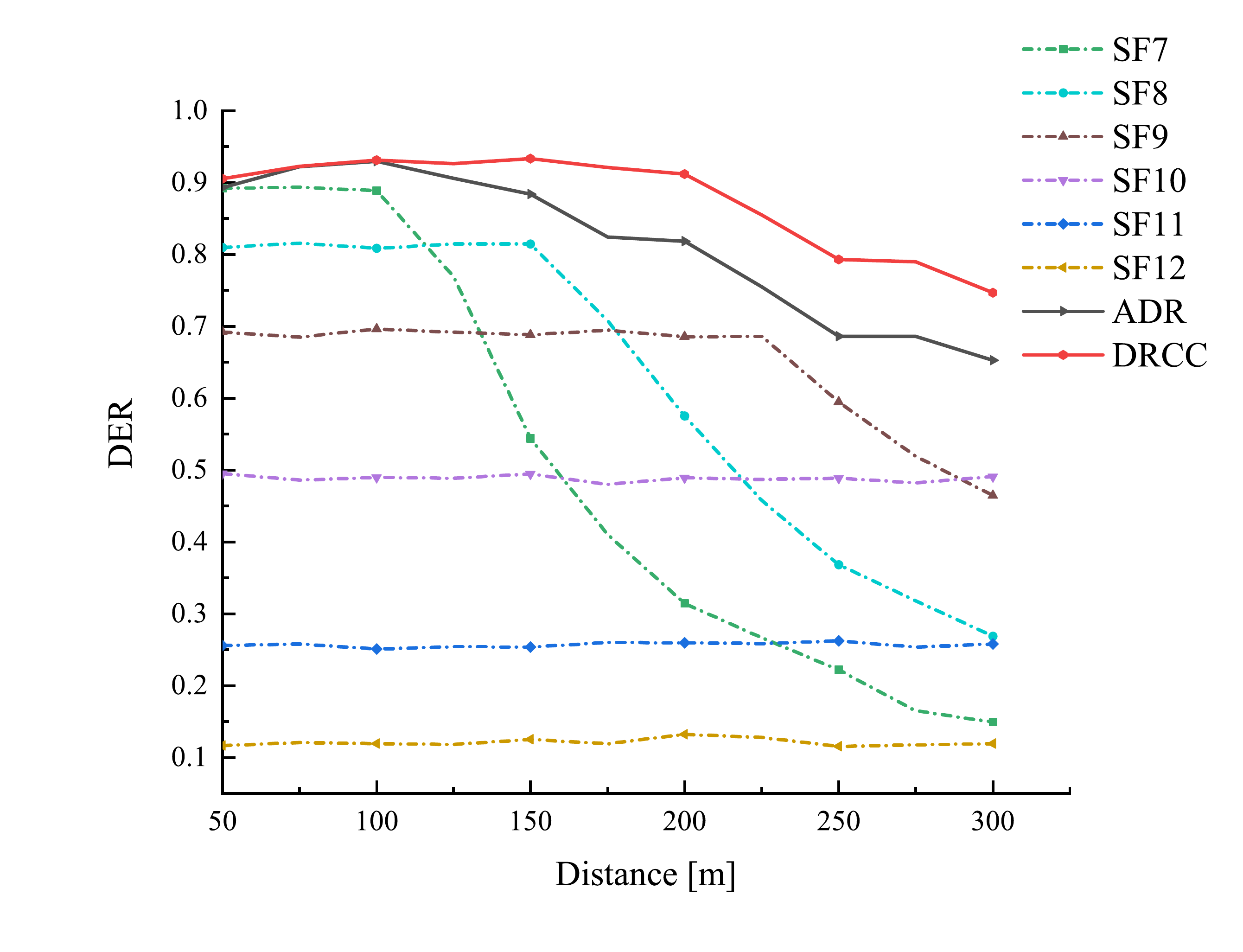}
	\caption{Impact of approaches on DER with the change in the scope of deployment.}
	\label{distance}
	\vspace{-1em}
\end{figure}

Fig.\ref{distance} shows the DER varies when the deployment scope changes. There are 1000 nodes distributed around a LoRa Gateway and these nodes send a packet every 100 seconds. For the methods using static communication parameters, the SF of each node is deterministic so that the maximum transmission distance is limited. As the scope of deployment continues to expand, some nodes may lose connect with the gateway when the RSSI is under the receiver sensitivity. Under the current path loss model, the maximum transmission distance is approximately 75/150/225 meters with SF=7/8/9 respectively. However, the curves of SF=10/11/12 change very slowly. It can be explained by that the communication range exceeds 300 meters. The DER is dominated by the collisions rather than the distance. ADR and DRCC performs well on the use of radio resources. Moreover, DRCC controls the channel selection of all nodes and balances the link load over all channels, which results in slightly better performance than ADR.

\begin{figure}[]
	\centering
	\includegraphics[width=0.49\textwidth]{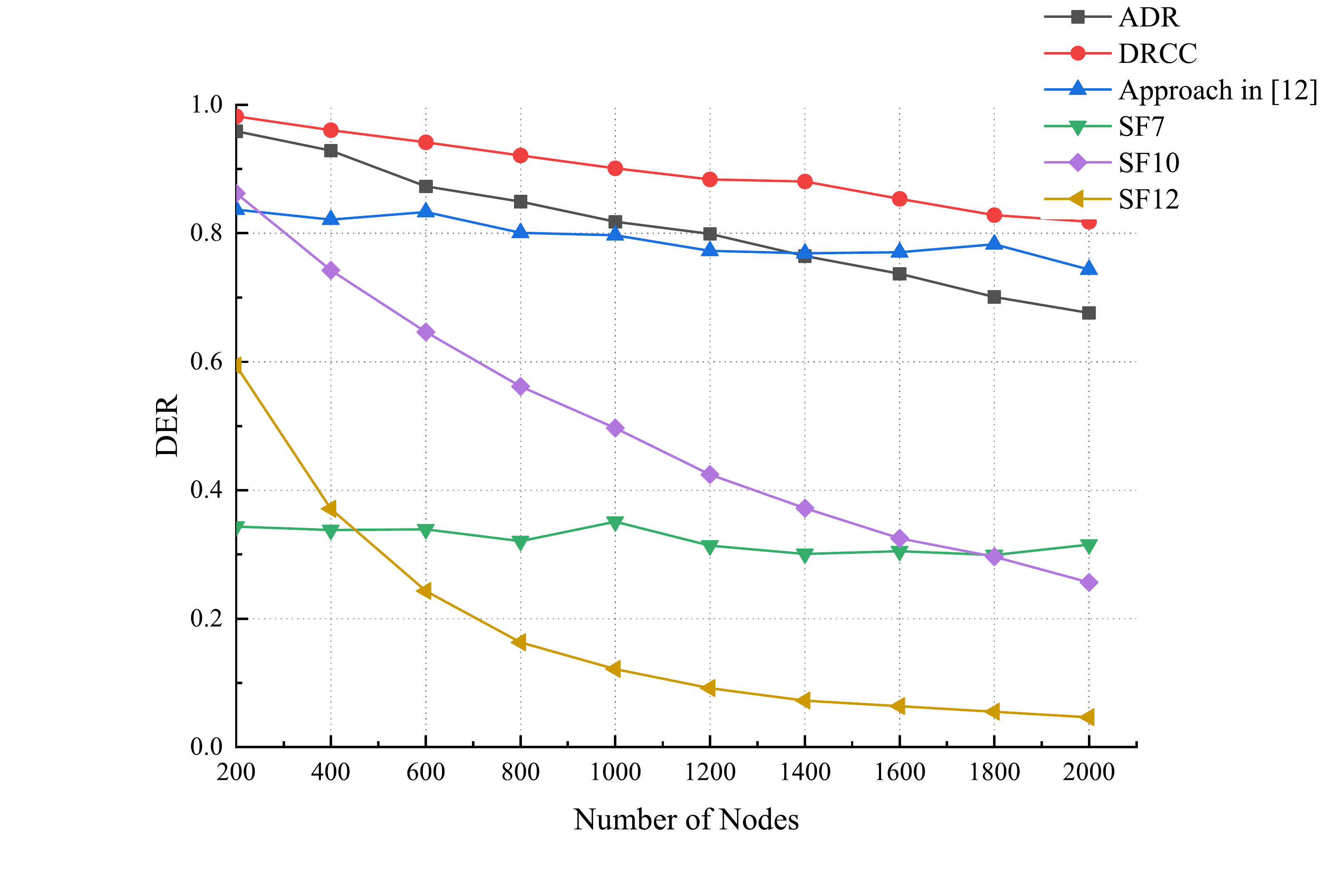}
	\caption{Impact of approaches on DER with different number of LoRa nodes.}
	\label{nodes}
	\vspace{-1em}
\end{figure}

The results of the third experiment are summarized in Fig.\ref{nodes}. The figure shows the varies of EDR with the number of nodes increases. All nodes are distributed in a circular range of 200 meters around a LoRa Gateway. The message rate is one packet sent every 100 seconds. The curve of SF7 has very little fluctuation since the maximum transmission distance using SF7 is less than 200 meters and a certain percentage of nodes cannot communicate with LoRa Gateway. With increasing the density of nodes in a cell, the curves of SF10 and SF12 drop sharply. There are quite a number of collisions due to massive LoRa nodes use the high SF. In addition, we also implement other three methods for comparison, \ie ADR, DRCC and the approach in \cite{power_sf_control}. As shown in Fig.\ref{nodes}, the performance of these approaches outperforms the static schemes, especially in dense deployments. It can be attributed to the SF orthogonality so that multiple spreading factors can be used to alleviate the interference between packets. The result of approach in \cite{power_sf_control} is stabilized at 0.8 and has minimally affected by increasing the number of nodes. DRCC can achieve significant performance improvements compared with ADR. If the DER is required to exceed 0.9 in the specific scenarios, using DRCC is able to support 1000 LoRa nodes while using ADR only supports 500 within a cell.

\section{Conclusion}
In the paper, we have proposed a novel data rate and channel control scheme named as DRCC for allocating wireless resources in LoRa networks. The main contributions lie in the combination of data rate allocation and channel selection control to improve network capacity. DRCC opportunistically adjusts the spreading factor based on the link-layer information of the short-term DER to cope with random loss. In addition, DRCC is able to balance the link load of all available channels so as to mitigate the access collision and increase the throughput. Finally, the experimental results demonstrate that DRCC outperforms other spreading factor allocation schemes in terms of reliability and performance under dense deployments.

\section*{Acknowledgment}
This work was supported by the National Natural Science Foundation of China (NSFC) under Grant Number 61671089.

\end{document}